\documentclass[epj]{webofc}

\def\xmm{\it XMM-Newton}

\def\igr{IGR J18245--2452}
\def\psr{PSR J1824--2452I}
\def\ltsima{$\; \buildrel < \over \sim \;$}
\def\simlt{\lower.5ex\hbox{\ltsima}}
\def\gtsima{$\; \buildrel > \over \sim \;$}
\def\simgt{\lower.5ex\hbox{\gtsima}}
\usepackage{graphicx}
\usepackage{psfig}
\usepackage{xspace}
\usepackage{paralist}
\usepackage{multicol}
\usepackage{url}

\usepackage[varg]{txfonts}   
\woctitle{Physics at the Magnetospheric Boundary}
\begin{document}
\title{Swinging between rotation and accretion power in a binary millisecond pulsar}
%
%

\author{A. Papitto\inst{1}\fnsep\thanks{\email{papitto@ice.csic.es}} \and C. Ferrigno \inst{2} \fnsep \and E. Bozzo \inst{2} \and N. Rea \inst{1,3}}

\institute{
Institute of Space Sciences (ICE; IEEC-CSIC), Campus UAB,
   Faculty of Science, Torre C5, parell, 2a planta, E-08193 Barcelona,
   Spain 
\and
ISDC Data Center for
     Astrophysics, Universit\'e de Gen\`eve, chemin d'\'Ecogia, 16, CH-1290
     Versoix, Switzerland
\and
Astronomical Institute ``Anton Pannekoek'', University of Amsterdam, Science Park 904, 1098 XH Amsterdam, The Netherlands
}

\abstract{%

We present the discovery of {\igr}, the first millisecond pulsar
observed to swing between a rotation-powered, radio pulsar state, and
an accretion-powered X-ray pulsar state \cite{papitto2013nature}. This
transitional source represents the most convincing proof of the
evolutionary link shared by accreting neutron stars in low mass X-ray
binaries, and radio millisecond pulsars. It demonstrates that swings
between these two states take place on the same time-scales of
luminosity variations of X-ray transients, and are therefore most
easily interpreted in terms of changes in the rate of mass in-flow.

While accreting mass, the X-ray emission of {\igr} varies dramatically
on time-scales ranging from a second to a few hours. We interpret a
state characterised by a lower flux and pulsed fraction, and by sudden
increases of the hardness of the X-ray emission, in terms of the onset
of a magnetospheric centrifugal inhibition of the accretion
flow. Prospects of finding new members of the newly established class
of transitional pulsars are also briefly discussed.

}
\maketitle
\section{Introduction}
\label{intro}

The extremely short spin periods of millisecond pulsars are assumed to
be the outcome of accretion of the mass transferred by a low mass
($\simlt$ M$_{\odot}$) companion star through an accretion disk
\citep{bisnovatyikogan1974}. This evolutionary framework is known as
the recycling scenario of old neutron stars (NS) in binaries
\citep{alpar1982,radhakrishnan1982}. After a Gyr-long mass accretion
phase during which the system appears as a bright X-ray source, the
mass transfer rate declines and allows the activation of a pulsar
powered by the rotation of its magnetic field, emitting from the radio
to the gamma-ray band.  The $\sim$300 millisecond radio pulsars (MSP)
in our Galaxy are consequently believed to be the recycled descendants
of accreting neutron stars in low mass X-ray binaries (NS-LMXB).  The
discovery of millisecond pulsations from accreting neutron stars in
LMXB, so called accreting millisecond pulsars
\citep{wijnands1998,chakrabarty1998,patruno2012arXiv}, proved how mass
accretion is able to effectively spin-up a neutron star to such fast
rotational velocities.

Millisecond pulsars in binary systems are also ideal laboratories to
study the interaction between the magnetosphere of a these quickly
spinning neutron star, and the plasma transferred by the donor
companion star. Their observational appearance is in fact assumed to
result from the balance between the outward pressure exerted by the
pulsar radiation, and the inward ram pressure of the in-falling
matter. Only when the pulsar emission manages to prevent the plasma to
enter inside the light cylinder of the pulsar (with radius
$R_{LC}\simeq 190$ km, for a NS spinning at 3.9 ms) a rotation-powered
pulsar is expected to turn on. On the contrary, when the in-falling
matter squeezes the magnetosphere inside the light cylinder, and
eventually reaches the neutron star surface, the system will appear as
a bright X-ray source. It was then proposed that transient NS-LMXB
would switch between rotation and accretion powered states as they
alternate between X-ray outbursts and quiescence
\citep{stella1994,campana1998,burderi2001}.


Here, we present the discovery of the first transitional system
observed to swing between rotation and accretion powered pulsar
states, definitely proving the evolutionary link shared by radio
millisecond pulsars and NS-LMXB \cite{papitto2013nature}.

\section{{\igr}, a transitional accreting millisecond pulsar}

\label{sec-1}

\begin{figure}[t!]
 \begin{center}
 \hbox{
\psfig{figure=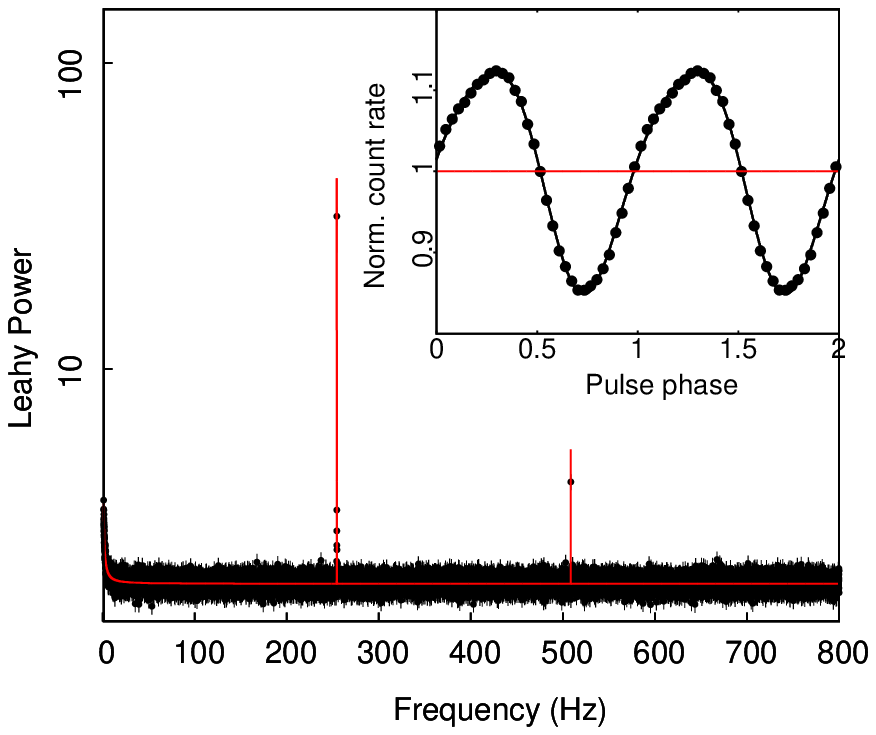,height=5.0cm,width=6.5cm}
 \hspace{0.0cm}
 \psfig{figure=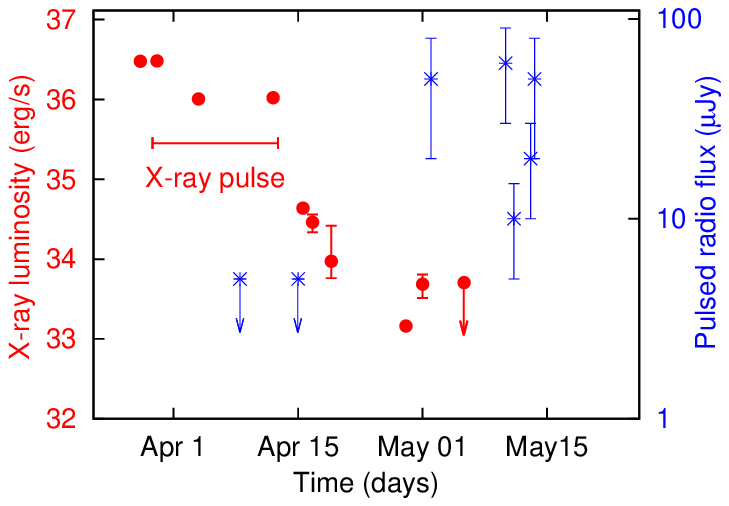,height=5.0cm,width=7.5cm}
 }
 \end{center}
 \caption{\footnotesize { {\it Left panel}: Leahy normalised power
     spectrum obtained averaging 128 s intervals of {\xmm} observations
     performed while {\igr} was in outburst on 13 April 2013. The
     inset shows the pulse profile accumulated in the 0.5--10 keV
     energy band. {\it Right panel}: observed X-ray luminosity (left scale, red
     dots) and pulsed radio flux (right scale, blue asterisks) of
     {\igr}/{\psr} during observations performed between 28 March and
     15 May 2013. The X-ray luminosity points refer to observations
     performed by INTEGRAL/ISGRI, Swift/XRT, XMM-Newton/EPICpn and
     Chandra, and are evaluated in the 0.5--10 keV band for a distance of
     5.5 kpc. Radio pulsed flux were evaluated at 2 GHz (Green Bank
     Radio Telescope) and 1.4 GHz (WSRT/Parkes). Only a subset of
     the available observations is shown, for illustrative purposes. See
     \cite{papitto2013nature} for the complete data set.  }}
 \label{fig-1}
 \end{figure}

{\igr} is a hard X-ray transient belonging to the globular cluster
M28, first detected in outburst by ISGRI/INTEGRAL on 28 March 2013
\citep{eckert2013}. At a distance of 5.5 kpc, the source attained an
X-ray luminosity of $L\approx3.5\times10^{36}$ erg s$^{-1}$ in both
soft (0.5-10 keV), and hard (20-100 keV) X-rays
\cite{eckert2013,romano2013}. Observations of thermonuclear explosions
caused by runaway burning of the material accreted onto the surface of
a neutron star \cite{papitto2013atel,linares2013} secured the
identification of the compact object in the system. Coherent
pulsations at a period of 3.9 ms were discovered in follow-up
XMM-Newton observations (see left panel of Fig.~\ref{fig-1}), making
this source the fifteenth accreting millisecond pulsar discovered so
far (the first by an observatory different than RXTE). The pulse
profile could be modelled with two harmonic components, with average
fractional amplitudes of 13.4(1) and 1.9(1) per cent, respectively
(see inset of left panel of Fig.~\ref{fig-1}). Analysis of the Doppler
shifts of the signal frequency revealed that the pulsar is in a 11.0
hr circular orbit around a $\approx0.2\,M_{\odot}$ companion star. The
X-ray luminosity, the thermonuclear bursts, and the presence of a
broad emission line at energies compatible with the K-$\alpha$
transition of iron, identify mass accretion as the driver of the
pulsed emission observed from {\igr}.

Cross-referencing with catalogues of radio pulsars \cite{atnf}, it was
realised that a source in M28, with the same spin and orbital
parameters, had been already observed as a rotation powered radio
pulsar a few years before, {\psr} \cite{begin2006}. This proved that
the source had performed a transition between the two states related
by the recycling scenario of neutron stars, representing its most
convincing observational evidence. The signal detected from this radio
pulsar was faint (few tenths of $\mu$Jy at 2 GHz) and irregularly
eclipsed by the presence of matter ejected by the pulsar wind.

The analysis of archival observations performed by Chandra revealed a
past X-ray brightening by an order of magnitude, above a quiescent
flux level of $10^{32}$ erg s$^{-1}$ \cite{papitto2013nature}. This is
a strong indication that also at that time the matter transferred by
the companion star started to spill out in the NS Roche lobe. However,
the source was in a much dimmer state than during the outburst in
2013, and it is possible that accretion onto the neutron star surface
was reduced or completely halted by magnetospheric centrifugal
inhibition (i.e. a propeller state, \cite{illarionov1975}).


While the 2008 X-ray brightening lagged by two months the most recent
previous detection of the radio pulsar, the reactivation of the radio
pulsar after the 2013 X-ray outburst was even more prompt
\cite{papitto2013nature}. An observational campaign performed by the
Green Bank, Parkes and Westerbork radio telescopes could detect a
pulsed radio signal, already a few weeks after the latest detection of
the X-ray pulsar, and a couple of days since the last detection in
X-rays (in which only an upper limit of 17 per cent of the pulse
amplitude could be set). To illustrate the fast swing between the two
states, we plot in the right panel of Fig.~\ref{fig-1} a sample of the
X-ray and radio observations of {\igr}/{\psr} during the months of
April and May 2013 (see \cite{papitto2013nature} for the complete
data-set).

The swings observed from {\igr} proved that transitions between
accretion and rotation powered states may take place on timescales
comparable to those characterising the alternation of X-ray outbursts
and quiescence in transient LMXB. When the system is found in X-ray
quiescence ($L_X\approx 10^{32}$ erg s$^{-1}$), the rate of mass
accretion onto the neutron star is lower than $10^{-14}$ M$_{\odot}$
yr$^{-1}$. The pressure exerted by the emission of a 3.9 ms
rotation-powered pulsar with a field of $10^8$--$10^9$ G (typical of
MSP) is able to keep the plasma beyond the light cylinder, and to
account for the observed X-ray luminosity. An increase of the mass
inflow rate may then push the magnetosphere inside the light cylinder,
switching off the radio pulsed emission. A bright ($L_X\approx
10^{36}$ erg s$^{-1}$) X-ray outburst is subsequently triggered after
enough mass has been loaded in the disk \cite{vanparadijs1996}, and
the magnetosphere is squeezed to a size of the order of the
co-rotation radius ($\approx 40$ km in {\igr}). Our observations
showed that as the mass accretion rate decreased at the end of the
outburst, a rotation-powered radio pulsar reactivated at most two
weeks after the last detection as an X-ray pulsar, and a couple of
days after the last detection in X-rays at a luminosity
($\mbox{few}\times10^{33}$ erg s$^{-1}$) above the quiescent level.

 \begin{figure}[t!]
 \begin{center}
 \hbox{
\psfig{figure=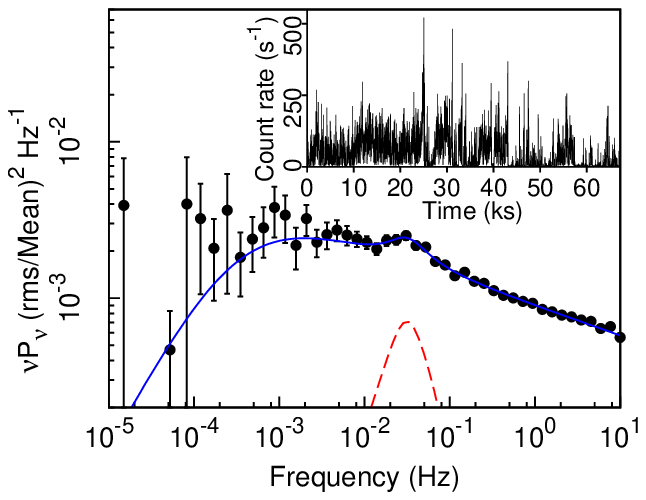,height=5.0cm,width=7cm}
 \hspace{0.0cm}
\psfig{figure=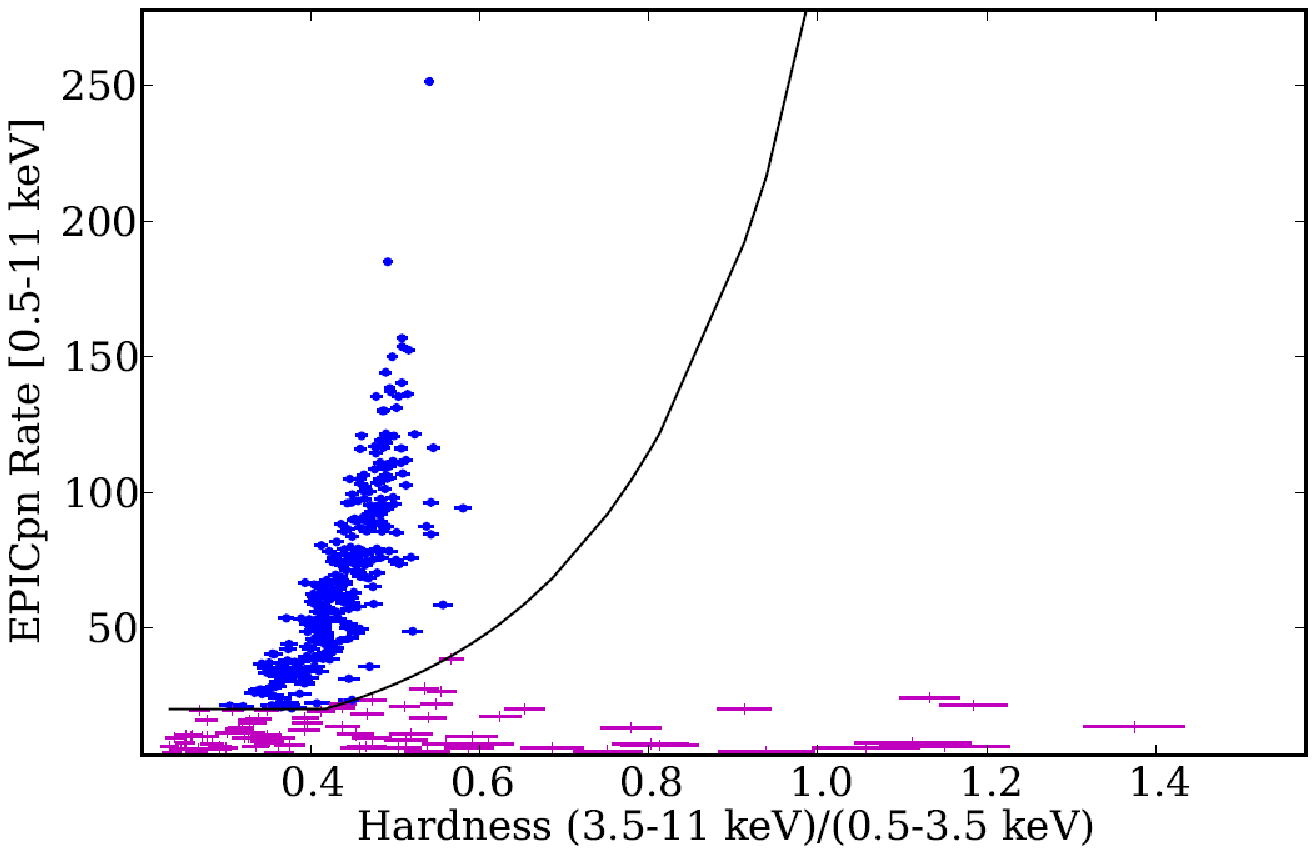,height=5.0cm,width=7cm}
 }
 \end{center}
 \caption{\footnotesize { {\it Left panel}: power density spectrum of
     a 67 ks XMM-Newton observation of {\igr} performed on 13 April
     2013, re-binned geometrically using a factor of 1.3, and
     subtracted of the white noise level. The inset shows the 0.5--10
     keV light curve.  {\it Right panel}: hardness-intensity 
     diagram of {\igr}. }}
 \label{fig-2}
 \end{figure}

\subsection{The X-ray variability}

During the X-ray outburst shown in 2013, {\igr} shows a marked and
peculiar variability of the magnitude and energy spectral distribution
of its X-ray emission (Ferrigno et al. 2013, A\&A, submitted; see the
inset of left panel of Fig.~\ref{fig-2} for the light curve observed
by XMM-Newton). A similar variability characterises also the radio
flux observed at 5.5 and 9 GHz by ATCA \cite{pavan2013}. The power
spectrum of the X-ray emission is dominated by a low frequency flicker
noise component, described by a power law, $P(f)\propto f^{-1.2}$,
cut-off at $\approx\mbox{few}\times 10^{-4}$ Hz (see left panel of
Fig.~\ref{fig-2}). Such a time-scale is compatible with the dynamical
time scale at the outer radius of a viscous disk in {\igr}, and
possibly indicates the propagation of fluctuations originating in the
whole disk down to the inner disk \cite{lyubarskii1997}, as a possible
explanation for the observed variability. This is at odds with the
properties of low frequency variability usually observed from
accreting millisecond pulsars, which show a noise with a cut-off at a
frequency in the 0.1--1 Hz range \cite{vanstraaten2005}.

The source follows two branches in a hardness-intensity diagram (see
right panel of Fig.~\ref{fig-2}). In the one marked by blue points,
the hardness varies in a relatively tight interval (0.3--0.5), in
correlation with the variations of the X-ray count rate between 30 and
250 counts per second. In the branch marked by magenta points, the
hardness performs swings up to $\approx1.4$ at a roughly constant, and
lower count rate level ($\sim 20$ counts s$^{-1}$). Considering that
the pulsed fraction is tightly correlated with the X-ray flux, thus
lower when the source is in the magenta branch, this low state can be
tentatively interpreted in terms of the onset of a centrifugal barrier
(i.e., a propeller state \cite{illarionov1975}). According to this
qualitative interpretation, the decrease of the X-ray flux and of the
pulsed fraction would be due to the (partial) inhibition of accretion
onto the neutron star surface, and in particular of the fraction of
matter falling close to the magnetic poles. The hardening of the
spectrum would be related to the less efficient cooling of a
Comptonizing cloud due to the lower flux of seed photons coming from
the NS surface, and/or to the energy deposited in the inner regions of
the disk by a propellering magnetosphere \cite{zhang1998}.

\section{Future prospects}


\begin{figure}
\centering
\includegraphics[width=5cm,angle=-90.0,clip]{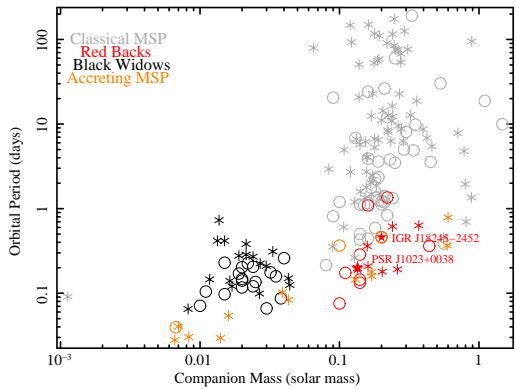}


\caption{Orbital period and minimum companion mass of field
     classical MSP with a white dwarf companion (grey), black widows
     (black), redbacks (red), and accreting millisecond pulsars
     (orange). Asterisks and circles mark sources in the Galactic
     field and in globular clusters, respectively.}
\label{fig-3}       
\end{figure}

{\igr} is the first member of a new class of pulsars, swinging between
rotation and accretion powered pulsar states, on timescales short
enough to make the transitions actually observable. More members of
this class should be searched among those binaries in which the
companion star currently fills its Roche lobe and spills out matter
towards the neutron star. The rate of mass transfer from the donor
star varies in fact over much longer timescales ($\simgt 50$ Myr;
e.g., \cite{tauris2012}).

The irregular eclipses of the pulsed emission observed from a number
of millisecond radio pulsars, as well as the variations of their
dispersion measure, are clear indications of pulsars whose radiation
is ejecting the matter that the companion tries to transfer. These
radio pulsars are usually dubbed as {\it black-widow} ($M_{c}<<0.1$
M$_{\odot}$; \cite{fruchter1988}) or {\it redbacks}
($M_{c}\sim0.2-0.4$ M$_{\odot}$; \cite{damico2001}), depending on the
mass of the companion star. Indeed, {\igr} is a redback, as well as
the other known transitional system, PSR J1023+0038, a radio pulsar
which showed optical emission lines in 2000-2001, indicating the
presence of an accretion disk at that time
\cite{archibald2009}\footnote{The system had an X-ray luminosity
  $\simlt10^{34}$ erg s$^{-1}$ at that time, so that accretion onto
  the neutron star surface was not taking place at a high rate, and
  the system was presumably in a propeller state.}.  This strongly
indicates that transitions to the accretion phase are expected to
happen among pulsars that are vaporising their companion. The number
of systems of this class is rapidly increasing during the last few
years \cite{roberts2013}, as they are bright gamma-ray emitters and
are easily detected by Fermi/LAT, increasing the probability of
finding new transitional systems in the near future.

On the other hand, during X-ray quiescence accreting millisecond
pulsars are mostly indistinguishable from rotation-powered pulsars, at
least for what concerns the irradiation of the companion star, and
their spin and orbital evolution
\cite{burderi2003,disalvo2008,hartman2008,patruno2010,papitto2011}. In
addition, accreting millisecond pulsars populate the same region of
the plot of orbital periods and companion star mass than {\it
  black-widow} and {\it redback} (see Fig.~\ref{fig-3}). However,
searches for a radio \cite{iacolina2009,iacolina2010} and gamma-ray
\cite{xing2013} have not been successful, so far. These non-detections
were possibly due to observational biases (e.g. the larger impact of
absorption of a radio signal by matter enshrouding the system, in
close binary systems) or to a paucity of gamma-ray photons. Future
observations with a longer exposure will be therefore crucial to
firmly assess the behaviour of these sources during quiescence.


\begin{acknowledgement}

Work done in the framework of the grants AYA2012-39303, as well as
SGR2009-811, and iLINK2011-0303. AP is supported by a Juan de la
Cierva Research Fellowship.

\end{acknowledgement}



%
\bibliography{nature.bib}
%
%
%
%

\end{document}